\documentclass[12pt]{article}

\usepackage[german,english]{babel}

\usepackage{amsmath}
\usepackage{amsfonts}
\usepackage{amsthm}

\newtheorem{theorem}{Theorem}[section]
\newtheorem{lemma}[theorem]{Lemma}

\theoremstyle{definition}
\newtheorem{define}[theorem]{Definition}
\newtheorem{assumption}[theorem]{Assumption}

\theoremstyle{remark}
\newtheorem{remark}[theorem]{Remark}

\numberwithin{equation}{section}

\begin{document}
\centerline{\LARGE\sc Stretched Exponential Relaxation}
\vspace{0.5 cm}
\centerline{\LARGE\sc in the Biased Random Voter Model}
\vspace{1 cm}
\centerline{Jan Naudts$^{\dagger,}$\footnote{E-mail: Jan.Naudts@uia.ua.ac.be.},
Frank Redig$^{\ddagger,}$\footnote{Postdoctoraal onderzoeker FWO.
E-mail: Frank.Redig@fys.kuleuven.ac.be.},
and Stefan Van Gulck$^{\dagger,}$\footnote{Bursaal IWT. E-mail:
Stefan.Vangulck@uia.ua.ac.be.}}
\vspace{0.2 cm}
\noindent
{\footnotesize $^\dagger$Departement Natuurkunde,
UIA,
Universiteitsplein 1,
B--2610 Antwerpen,
Belgium}
\\
{\footnotesize $^\ddagger$Instituut voor Theoretische Fysica,
KUL,
Celestijnenlaan 200D,
B--3001 Leuven,
Belgium}
\begin{abstract}
\noindent
We study the relaxation properties of the voter model with i.i.d.\ random bias.
We prove under mild condions that the disorder-averaged relaxation of this
biased random voter model is faster than a stretched exponential with exponent
$d/(d+\alpha)$, where $0<\alpha\le2$ depends on the transition rates of the
non-biased voter model.
Under an additional assumption, we show that the above upper bound is optimal.
The main ingredient of our proof is a result of Donsker and Varadhan (1979).
\end{abstract}


\section{Introduction}

Recently, important progress has been made in the rigorous analysis of the
dynamical properties of random spin systems.
The best known class of random spin systems consists of spin glasses equiped
with a Glauber dynamics (kinetic Ising models with random interaction and
single spin-flip time evolution).
These stochastic models are used in Monte Carlo simulations to mimic the time
evolution of disordered systems, see e.g.\ Ogielski (1985).

The following results are stated with probability one with respect to the
disorder and uniform in the initial condition.
The assumptions under which they are derived differ slightly
(cf.\ the original papers for the precise statements).
Zegarlinski (1994) proves the absence of a spectral gap in
random Glauber models in dimension $d\ge1$.
As a consequence, the decay of time dependent correlations in equilibrium cannot
be exponentially fast.
Moreover, he shows in dimension $d=1$ that the decay of local
functions can be bounded above by a stretched exponential whose exponent can be
chosen arbitrarily in the open interval $(0,1)$.
Guionnet and Zegarlinski (1996) extend Zegarlinski's result.
In particular, they find the same stretched exponential bound in dimension
$d=2$, even for continuous-spin systems.
For $d\ge3$ they are able to derive a relaxation faster than algebraic
(Guionnet and Zegarlinski, 1997).
Cesi, Maes and Martinelli (1997a) succeed in establishing for the Glauber
model in arbitrary dimension $d\ge1$, but under more restricted conditions on
the distribution of the interaction, an upper bound for the asymptotic
relaxation of local observables with decay faster than any stretched
exponential.
Furthermore, they obtain under even more severe assumptions (not valid for the
diluted Ising model) a lower bound which is similar to the upper bound, implying
that the latter cannot be improved in general.

Cesi, Maes and Martinelli (1997a) study also the (physically more relevant)
disorder-averaged relaxation of Glauber models.
In dimension $d\ge2$ they find an upper bound with decay slower than any
stretched exponential but faster than algebraic.
In a particular case, they derive a lower bound which is again similar to the
upper bound.
Both bounds are in agreement with a non-rigorous result for the asymptotic
relaxation of the spin autocorrelation function in the diluted Ising model below
the treshold for bond percolation; see Bray (1989), and references therein.

Alexander, Cesi, Chayes, Maes and Martinelli (1998) confirm for the
diluted Ising model in dimension $d\ge2$ all the results of Cesi, Maes and
Martinelli (1997a), even above the percolation treshold, except the almost sure
lower bound which they prove only with positive probability; see also Cesi, Maes
and Martinelli (1997b), and Martinelli (1997).

The previously mentioned results concern spin systems whose single-flip
dynamics are reversible with respect to some Gibbs measure corresponding to a
random potential.
A powerful tool in the analysis of these systems is the so-called logarithmic
Sobolev inequality.
Gielis and Maes (1996) tackle the problem of estimating the asymptotic
relaxation in a more general class of random spin systems with quenched
disorder.
Their method consists of a coupling of the spin system to the contact process
and the use of percolation techniques on the graphical representation of the
contact process.
They obtain, with probability one with respect to the disorder, an upper bound
which decays faster in time than any power law; see Remark~\ref{sec-gm96} below.
Moreover, in case of directed interactions they improve the upper bound to a
stretched exponential whose exponent can be taken arbitrary in the open interval
$(0,1/2)$.
No lower bound is known.
However, it is believed that their estimate is far from optimal.
In a forthcoming paper we will ameliorate their bound for the random voter
model with quenched bias.

To the best of our knowledge, no rigorous bounds are yet available for the
disorder-averaged relaxation of spin systems in this larger class.
In the present paper we prove, under the mild assumption of a positive
probability for each voter to be biased, that the disorder-averaged
asymptotic relaxation of local functions in the biased random voter model is,
uniform in the initial condition, bounded from above by a stretched exponential
decay in time with exponent $d/(d+\alpha)$, where $d\ge1$ is the dimension of
the hypercubic lattice $\mathbb{Z}^d$ and $0<\alpha\le2$ depends only on the
transition rates of the non-biased voter model.
Moreover, the upper bound cannot be improved in general:
Under the assumption that each voter is non-biased with positive probability, we
prove that the disorder-averaged asymptotic relaxation of monotone local
functions is bounded from below by a stretched exponential with the same
exponent.
The approach of our proof consists of a coupling of the dual process of the
biased voter model with a collection of independent random walks, and the
application of a result of Donsker and Varadhan (1979) about the number of
distinct sites visited by a random walk.

The model studied in this paper is an example of a spin system with an annealed
random field.
Often a distinction is made between spin systems with a random field and those
with a random coupling between the constituents.
The most famous example of the latter kind is the kinetic Ising model with
an exchange interaction of random strength
(as e.g.\ in the diluted Ising model).

We have restricted the above discussion to rigorous results for random spin
systems.
There are other particle systems in random environment worth mentioning.
An extensively studied class of such systems consists of random walks in random
environment; see e.g.\ den Hollander (1984), den Hollander, Naudts and
Scheunders (1992), and den Hollander, Naudts and Redig (1992, 1994).
The random trap model has even been studied with a time-evolving
environment in den Hollander and Shuler (1992), and Redig (1994).
The asymptotics of the survival probability in the random trap model is closely
related to our result; see the discussion in den Hollander (1984), and the
references given there.

The physics behind the phenomenon of slower than exponential decay is in all
these cases the same. In the diffusion model with traps there exists large
regions free of traps, be it with a small probability. Particles diffusing
in trap-free regions will survive significantly longer than those in areas
with average or high density of traps, contributing in an anomalous way to
the asymptotics of the decay. Similarly, in spin glasses the ordering of the
spins decays very slowly in areas of strong interaction. In the present
model areas with a small amount of biased voters can display
a dominancy of opinion 1 during a long time, before decaying to opinion 0.

This paper is organized as follows. In Section~\ref{sec-result} we define the
biased voter model and its randomized version, state our result, and mention the
theorem of Donsker and Varadhan (1979) on which the proof of our result relies
heavily. In Section~\ref{sec-proof} we describe the coalescing dual process of
the biased voter model, and proof the upper and lower bound in
Theorem~\ref{sec-main}.

\section{Statement of result}
\label{sec-result}

A spin system is a continuous-time Markov process $(\eta_t)_{t\ge0}$ with
configuration space
\begin{equation}
\mathbb{X}=\{0,1\}^{\mathbb{Z}^d}=
\bigl\{\eta:\mathbb{Z}^d\rightarrow\{0,1\}:x\mapsto\eta(x)\bigr\}.
\end{equation}
Let $\eta^x$ be the configuration obtained from configuration $\eta$ by changing
the state at site $x$, i.e.,
\begin{equation}
\eta^x(y)
=
\begin{cases}
1-\eta(x)
&\text{if $y=x$,}
\\
\eta(y)
&\text{if $y\not=x$.}
\end{cases}
\end{equation}
The infinitesimal dynamics of a spin system is specified by a Markov
pregenerator $\Omega$, given by
\begin{equation}
\label{eq:generator}
\Omega f(\eta)=\sum_{x\in\mathbb{Z}^d}c(x,\eta)[f(\eta^x)-f(\eta)]
\end{equation}
and defined on a suitable subclass of $C(\mathbb{X})$, the set of real-valued
continuous functions with domain $\mathbb{X}$.
(On $\{0,1\}$ we assume the discrete topology and on $\mathbb{X}$ the product
topology.)
Such a suitable subclass is the collection of local functions defined below, see
Liggett (1985).
The non-negative (and uniformly bounded) quantities $c(x,\eta)$ in
\eqref{eq:generator} are called the
transition rates of the spin system.
We start by specifying the transition rates of the biased (random) voter model.

\begin{define}
The {\em biased voter model\/} is the spin system with transition rates
\begin{equation}
\label{eq:bvrates}
c_\beta(x,\eta)
=
\beta(x)\eta(x)
+
\sum_{y\in\mathbb{Z}^d}p(y-x)\{\eta(x)[1-\eta(y)]+\eta(y)[1-\eta(x)]\}
\end{equation}
where
\begin{subequations}
\begin{gather}
\beta:\mathbb{Z}^d\rightarrow[0,\infty),
\quad
\sup_{x\in\mathbb{Z}^d}\beta(x)<\infty,
\\
p:\mathbb{Z}^d\rightarrow[0,1],
\quad
\sum_{x\in\mathbb{Z}^d}p(x)=1,
\quad
\text{and}
\quad
p(0)=0.
\end{gather}
\end{subequations}
The {\em biased random voter model\/} is the spin system with transition rates
\eqref{eq:bvrates} where $\{\beta(x),x\in\mathbb{Z}^d\}$ is a collection of
independent identically distributed random variables with joint distribution
$\mathbb{B}$.
\end{define}

The interpretation of the non-biased voter model (i.e., the spin system with
transition rates \eqref{eq:bvrates}, but with $\beta=0$) is well known, see
e.g.\ Liggett (1985).
In this interpretation $\mathbb{Z}^d$ is viewed as a collection of individuals,
each of which has one of two possible opinions (0 or 1) on a political issue.
At the event times of a Poisson process the individual at site $x$ reassesses
his opinion $\eta(x)$ in the following way: he consults a ``friend'' $y$ with
probability $p(y-x)$ and then adopts his position $\eta(y)$.
The presence of $\beta$ in the transition rates of the biased voter model can be
interpreted as the presence of a propaganda mechanism which drives each
individual to opinion 0; $\beta(x)$  is then a measure for the susceptibility of
individual $x$ for that propaganda.
Other interpretations of the biased (random) voter model are possible. E.g., it
can be viewed as a zero-temperature kinetic Ising model in a (random) magnetic
field.

From the explicit form of the transition rates it is immediate that
configuration $\eta=0$ is absorbing, and thus $\delta_0$, the Dirac measure
with unit mass at $\eta=0$, is invariant.
From the well-known ergodicity criterion $M<\varepsilon$ (Liggett, 1985) or from
duality (see below) one can deduce that the biased voter model is
(exponentially) ergodic whenever $\beta$ is uniformly positive, i.e.
\begin{equation}
\inf_{x\in\mathbb{Z}^d}\beta(x)>0,
\end{equation}
irrespective of the values of $p(x)$.
On the other hand, the non-biased voter model ($\beta=0$) is not ergodic.
A more refined ergodicity criterion for the biased voter model should, of
course, contain conditions on $p(x)$, e.g.\ irreducibility.
\\[0.5 cm]
For the formulation of our main result we need the following definitions and
assumption.

\begin{define}
On $\mathbb{X}$ we define the partial order relation $\le$ by
\begin{equation}
\eta\le\zeta
\Leftrightarrow
\eta(x)\le\zeta(x)
\text{ for every }
x\in\mathbb{Z}^d.
\end{equation}
A real-valued function $f:\mathbb{X}\rightarrow\mathbb{R}$ is said to be
monotone whenever
\begin{equation}
\eta\le\zeta
\Rightarrow
f(\eta)\le f(\zeta).
\end{equation}
A spin system with transition rates $c(x,\eta)$ is said to be attractive if
\begin{equation}
\eta\le\zeta
\Rightarrow
\begin{cases}
c(x,\eta)\le c(x,\zeta)
&\text{if $\eta(x)=\zeta(x)=0$}
\\
c(x,\eta)\ge c(x,\zeta)
&\text{if $\eta(x)=\zeta(x)=1$}
\end{cases}
\end{equation}
Thus, a spin system is attractive if, for every $x\in\mathbb{Z}^d$,
$[1-2\eta(x)]c(x,\eta)$ is a monote function of $\eta$.
\end{define}

\noindent
It is well known that a spin system is attractive if and only if it is
monotone, in the sense that $S(t)f$, defined by $S(t)f(\eta)=\mathbb{E}^\eta
f(\eta_t)$, is monotone if $f$ is monotone
(Liggett, 1985).

\begin{define}
\label{sec-local}
Let $\mathbb{Y}=\{A: A \text{ finite subset of } \mathbb{Z}^d\}$.
This is a countable set.
A function $f:\mathbb{X}\rightarrow\mathbb{R}$ is said to be local if there
exists an $A\in\mathbb{Y}$ such that
\begin{equation}
\label{eq:local}
\forall \eta,\zeta\in\mathbb{X}:
\eta=\zeta \text{ on } A\Rightarrow f(\eta)=f(\zeta),
\end{equation}
where $\eta=\zeta$ on $A$ means that $\eta(x)=\zeta(x)$ for every $x\in A$.
The smallest set $A$ for which \eqref{eq:local} holds will be denoted by
$\Lambda(f)$.
The space of continuous $C(\mathbb{X})$ will be equiped with the uniform norm
$||\cdot||$, defined by
\begin{equation}
||f||=\sup_{\eta\in\mathbb{X}}|f(\eta)|.
\end{equation}
With respect to this norm, the set of local functions is dense in
$C(\mathbb{X})$ (Stone--Weierstrass).
\end{define}

\noindent
In the statement of Theorem 1, and Theorem 2, we will suppose that the following
key assumption holds.

\begin{assumption}
\label{sec-assume}
Let, for every $k\in\mathbb{R}^d$,
\begin{equation}
\Hat{p}(k)=\sum_{x\in\mathbb{Z}^d}p(x)\exp(i\langle k,x\rangle),
\end{equation}
where $\langle\cdot,\cdot\rangle$ is the Euclidean scalar product defined on
$\mathbb{R}^d$.
Assume that
\begin{itemize}
\item $\Hat{p}(k)=1$ if and only if $k=2\pi(n_1,\ldots,n_d)$, where
      $n_1,\ldots,n_d$ are integers.
\item $\Hat{p}(k)=1-D(k)+o(|k|^\alpha)$ as
      $|k|=\sqrt{\langle k,k\rangle}\rightarrow0$, where $0<\alpha\le2$ and
      $\exp(-D(k))$ is the characteristic function of a symmetric stable law of
      index $\alpha$ in $\mathbb{R}^d$ which is non-degenerate, i.e.,
      \begin{equation}
      D(k)=
      \begin{cases}
      \displaystyle
      \sum_{i,j=1}^d D_{ij}k_ik_j
      \text{ for some positive definite matrix $D_{ij}$,}
      &\text{if $\alpha=2$,}
      \\
      \displaystyle
      \int_{\mathbb{S}^d}\int_0^\infty[1-\cos\langle k,ry\rangle]\,
      r^{-(1+\alpha)}\,dr\,\sigma(dy),
      &\text{if $\alpha<2$.}
      \end{cases}
      \end{equation}
      In the expression for $D(k)$, $\sigma$ is a symmetric measure on the unit
      ball $\mathbb{S}^d$ in $\mathbb{R}^d$, and the assumption of
      non-degeneracy means that the support of $\sigma$ spans $\mathbb{S}^d$.
\end{itemize}
The above assumptions on $\Hat{p}$ imply that the random walk with transition
probabilities $p(x)$ is irreducible and the distribution $p(x)$ belongs to
the domain of normal attraction of a non-degenerate symmetric stable law of
index $0<\alpha\le2$ (Spitzer, 1976).
\end{assumption}

\begin{theorem}[main result]
\label{sec-main}
Suppose that Assumption~\ref{sec-assume} holds.
\begin{itemize}
\item[\rm({\em i\/})]  If\/ $\mathbb{B}(\{\beta:\beta(0)\not=0\})>0$, then
                       there exists a constant $\nu_1>0$ such that
                       \begin{subequations}
                       \label{eq:bounds}
                       \begin{equation}
                       \label{eq:upper}
                       \limsup_{t\rightarrow\infty}t^{-d/(d+\alpha)}
                       \log\int||S_\beta(t)f-\delta_0(f)||\,\mathbb{B}(d\beta)
                       \le
                       -C_{d,\alpha}(\nu_1)
                       \end{equation}
                       for every local function $f$.
\item[\rm({\em ii\/})] If\/ $\mathbb{B}(\{\beta:\beta(0)=0\})>0$, then there
                       exists a constant $\nu_2>\nu_1$ such that
                       \begin{equation}
                       \label{eq:lower}
                       \liminf_{t\rightarrow\infty}t^{-d/(d+\alpha)}
                       \log\int||S_\beta(t)f-\delta_0(f)||\,\mathbb{B}(d\beta)
                       \ge
                       -C_{d,\alpha}(\nu_2)|\Lambda(f)|
                       \end{equation}
                       \end{subequations}
                       for every non-constant monotone local function $f$.
\end{itemize}
In (\ref{eq:bounds}) $C_{d,\alpha}$ is an increasing function defined on\/
$[0,\infty)$ given by
\begin{equation}
C_{d,\alpha}(\nu)=
(d+\alpha)\Bigl[\Bigl(\frac{\lambda}{d}\Bigr)^d
\Bigl(\frac{\nu}{\alpha}\Bigr)^\alpha\Bigr]^{1/(d+\alpha)}
\end{equation}
where $\lambda>0$ is a constant specified in Theorem~\ref{sec-dovar} below.
\end{theorem}

\begin{remark}
\label{sec-constants}
In the proof of Theorem~\ref{sec-main} we will choose the following values for
$\nu_1$ and $\nu_2$:
\begin{equation}
\begin{split}
\nu_1
&=
-\log\int\frac{1}{1+\beta(0)}\,\mathbb{B}(d\beta)
\\
\nu_2
&=
-\log\mathbb{B}(\{\beta:\beta(0)=0\})
\end{split}
\end{equation}
Using Jensen's inequality one sees easily that
$\nu_1\le\log(1+\Bar{\beta})$, where
$\Bar{\beta}=\int\beta(0)\,\mathbb{B}(d\beta)$.
\end{remark}

\begin{remark}
\label{sec-gm96}
The relaxation for typical realizations of the disorder in a general class of
(quenched) random spin systems has been studied by Gielis and Maes (1996).
Their result for the biased random voter model reads in our notations as
follows; see also Klein (1994).
{\em Suppose $p(x)=0$ whenever $|x|\not=1$, and let
$K>\bigl(d+\sqrt{d(d+1)}\bigr)^2$.
Then, there exists a constant $v_0=v_0(K,d)>1$ such that for all $1<v<v_0$ and
$m>0$ there is a positive constant $\varepsilon=\varepsilon(K,d,m,v)$ such that}
\begin{equation}
\label{eq:gmcon}
\int\Bigl[\log\Bigl(1+\frac{2}{\beta(0)}\Bigr)\Bigr]^K\,\mathbb{B}(d\beta)
<
\varepsilon
\end{equation}
{\em implies}
\begin{equation}
\limsup_{t\rightarrow\infty}
\frac{||S_\beta(t)f-\delta_0(f)||}{\bigl(\log(1+t)\bigr)^v}
\le-m
\end{equation}
{\em for every local function $f$ and\/ $\mathbb{B}$-almost every realization
$\beta$ of the disorder.\/}
Notice that a necessary condition for \eqref{eq:gmcon} to hold is
$\mathbb{B}(\{\beta:\beta(0)=0\})=0$.
\end{remark}

\noindent
Theorem~\ref{sec-main} will be proven in Section~\ref{sec-proof}.
The main ingredient of the proof is the following result of Donsker and
Varadhan (1979).
The restrictions we pose on the stochastic matrix $p(x,y)$ in
Assumption~\ref{sec-assume} are sufficient for the validity of this result.

\begin{theorem}[Donsker and Varadhan]
\label{sec-dovar}
Let $R_t$ be the colection of (distict) sites visited in the time interval\/
$[0,t]$ by a continuous-time random walk $(X_t)_{t\ge0}$ on $\mathbb{Z}^d$ with
transition probabilities\/ $\mathbb{P}^x(X_t=y)=p_t(y-x)$, where
\begin{equation}
p_t(x)=e^{-t}\sum_{n=0}^\infty\frac{t^n}{n!}p^{(n)}(x)
\end{equation}
and $p^{(n)}(x)$ are the $n$-step transition probabilities of the
discrete-time random walk with one-step transition probabilities $p(x)$
(Spitzer, 1976).
Suppose $L$ is the infinitesimal generator of the symmetric stable process in
$\mathbb{R}^d$ of index $0<\alpha\le2$ corresponding to $D(k)$, i.e.,
\begin{equation}
Lf(x)
=
\begin{cases}
\displaystyle
\sum_{i,j=1}^d D_{ij}\frac{\partial^2}{\partial x_i\partial x_j}f(x)
&\text{if\/ $\alpha=2$}
\\
\displaystyle
\int_{\mathbb{S}^d}\int_0^\infty\Bigl[\frac{f(x+ry)+f(x-ry)}{2}-f(x)\Bigr]
\,r^{-(1+\alpha)}\,dr\,\sigma(dy)
&\text{if\/ $\alpha<2$}
\end{cases}
\end{equation}
for every smooth real-valued function $f$ on $\mathbb{R}^d$.
Then, for every $\nu>0$,
\begin{equation}
\lim_{t\rightarrow\infty}t^{-d/(d+\alpha)}\log\mathbb{E}^0\exp(-\nu|R_t|)
=
-C_{d,\alpha}(\nu),
\end{equation}
where $C_{d,\alpha}(\nu)$ is defined in Theorem~\ref{sec-main} and
\begin{equation}
\lambda=\inf_G\lambda(G)>0,
\end{equation}
the infimum being over all open subsets $G$ in $\mathbb{R}^d$ of unit volume and
$\lambda(G)$ is the smallest eigenvalue of $-L$ with Dirichlet boundary
conditions for $G$.
\end{theorem}

\section{Proof of Theorem~\ref{sec-main}}
\label{sec-proof}

The proof of Theorem~\ref{sec-main} is an application of
Theorem~\ref{sec-dovar}, combined with attractiveness and duality.
We start with a description of the coalescing dual process of the biased voter
model (see also Liggett, 1985).

\subsection{Duality for the biased voter model}

The dual process is a continuous-time Markov chain with configuration space
$\mathbb{Y}$, see Definition~\ref{sec-local} and the construction below.
In case of coalescing duality the choice for the duality function is
\begin{equation}
\label{eq:dualfun}
H:\mathbb{X}\times\mathbb{Y}\rightarrow\{0,1\}:
(\eta,A)\mapsto\prod_{x\in A}\eta(x)
\end{equation}
In the proof of Theorem~\ref{sec-main} we will make use of the property that
$H(\eta,A)$ is increasing in the first argument and decreasing in the second.
Applying operator $\Omega_\beta$, defined in \eqref{eq:generator} and
\eqref{eq:bvrates}, to the local
function $H(\cdot,A)$ gives, for $A\in\mathbb{Y}$,
\begin{equation}
\begin{split}
\Omega_\beta H(\eta,A)
&=
\sum_{x\in\mathbb{Z}^d}c_\beta(x,\eta)[H(\eta^x,A)-H(\eta,A)]
\\
&=
\sum_{B\in\mathbb{Y}}q(A,B)[H(\eta,B)-H(\eta,A)]-V_\beta(A)H(\eta,A)
\end{split}
\end{equation}
where
\begin{equation}
V_\beta(A)=\sum_{x\in A}\beta(x)
\quad\text{(Feynman--Kac potential)}
\end{equation}
and
\begin{equation}
q(A,B)
=
\begin{cases}
\sum_{y\in A}p(y-x)
&\text{if $\exists\,x\in A:B=A\setminus\{x\}$}
\\
p(y-x)
&\text{if $\exists\,x\in A,\exists\,y\not\in A:B=(A\setminus\{x\})\cup\{y\}$}
\\
0
&\text{otherwise}
\end{cases}
\end{equation}
The $q(A,B)$ are the transition rates of a continuous-time Markov chain
$(A_t)_{t\ge0}$ on $\mathbb{Y}$, which is called the coalescing dual
process of the biased voter model.
This terminology finds its origin in the interpretation of $A_t$ as the finite
collection of occupied sites in $\mathbb{Z}^d$ at time $t$\/; each site is
occupied by at most one particle and only a finite number of particles are
present.
The transition rates $q(A,B)$ show that a particle at site $x\in\mathbb{Z}^d$ is
removed after an exponentially distributed waiting time with mean 1, independent
of the attendance of particles at other sites, and replaced with probability
$p(y-x)$ by a particle at site $y\in\mathbb{Z}^d$.
If after this action two particles occupy the same site $y$, these two particles
coalesce.

Notice that the transition rates $q(A,B)$ are independent of the bias $\beta$
and, in particular, the coalescing dual processes of the biased and the
non-biased voter model coincide.

The following duality relation (or Feynman--Kac formula) explains why
$(A_t)_{t\ge0}$ is called a dual process of the biased voter model:
\begin{equation}
\label{eq:dualrel}
\mathbb{E}^\eta_\beta H(\eta_t,A)
=
\mathbb{E}^A H(\eta,A_t)\exp\Bigl\{-\int_0^t V_\beta(A_s)\,ds\Bigr\}
\end{equation}
A proof of equation \eqref{eq:dualrel} can be found in Liggett (1985).
Remark that the bias introduces a Feynman--Kac term in \eqref{eq:dualrel}.

In the proof of Theorem~\ref{sec-main} we will use the property that every local
function $f$ has a unique representation of the form
\begin{equation}
\label{eq:localrep}
f(\eta)=\sum_{A\in\mathbb{Y}}\Hat{f}(A)H(\eta,A).
\end{equation}
Obviously, $\Hat{f}(A)=0$ unless $A\subset\Lambda(f)$.

\subsection{Upper bound}

In this section we use the notation
\begin{equation}
\Sigma(f)=\sum_{A\not=\emptyset}|\Hat{f}(A)|,
\end{equation}
where $f$ is an arbitrary local function,
and start with the following estimate:
\begin{equation}
\label{eq:estim}
\begin{split}
||S_\beta(t)f-\delta_0(f)||
&\le
\sum_{A\not=\emptyset}|\Hat{f}(A)|\sup_{\eta\in\mathbb{X}}\mathbb{E}_\beta^\eta
H(\eta_t,A)
\\
&=
\sum_{A\not=\emptyset}|\Hat{f}(A)|\mathbb{E}_\beta^{\eta=1}H(\eta_t,A)
\\
&\le
\Sigma(f)\max_{x\in\Lambda(f)}\mathbb{E}^1_\beta H(\eta_t,\{x\})
\\
&=
\Sigma(f)\max_{x\in\Lambda(f)}\mathbb{E}^{\{x\}}
\exp\Bigl\{-\int_0^t V_\beta(A_s)\,ds\Bigr\}
\\
&=
\Sigma(f)
\max_{x\in\Lambda(f)}\mathbb{E}^x\exp\Bigl\{-\int_0^t \beta(X_s)\,ds\Bigr\}
\end{split}
\end{equation}
The first step in \eqref{eq:estim} follows from identity
\eqref{eq:localrep} and
\begin{equation}
\int H(\eta,A)\,\delta_0(d\eta)
=
\begin{cases}
1
&\text{if $A=\emptyset$}
\\
0
&\text{if $A\not=\emptyset$}
\end{cases}.
\end{equation}
The second step in \eqref{eq:estim} uses the attractiveness of the
biased voter model and the property that the duality function is increasing in
its first argument.
The third step in \eqref{eq:estim} is a consequence of the fact that the
duality function is decreasing in its second argument.
The fourth step in \eqref{eq:estim} is a particular case of the duality
relation \eqref{eq:dualrel}, while the fifth step uses that the dual process
$(A_t)_{t\ge0}$ with initial configuration $\{x\}$ is identical to the random
walk $(X_t)_{t\ge0}$ starting from position $x$.
Integrating \eqref{eq:estim} with respect to $\mathbb{B}$ we get
\begin{equation}
\label{eq:Bavestim}
\begin{split}
\int||S_\beta(t)f-\delta_0(f)||\,\mathbb{B}(d\beta)
&\le
\Sigma(f)\int\max_{x\in\Lambda(f)}
\mathbb{E}^x\exp\Bigl\{-\int_0^t\beta(X_s)\,ds\Bigr\}\,\mathbb{B}(d\beta)
\\
&\le
\Sigma(f)\int\sum_{x\in\Lambda(f)}
\mathbb{E}^x\exp\Bigl\{-\int_0^t\beta(X_s)\,ds\Bigr\}\,\mathbb{B}(d\beta)
\\
&=
\Sigma(f)|\Lambda(f)|
\,\mathbb{E}^0\int\exp\Bigl\{-\int_0^t\beta(X_s)\,ds\Bigr\}\,\mathbb{B}(d\beta)
\end{split}
\end{equation}
The last step in \eqref{eq:Bavestim} uses Fubini's theorem and the translation
invariance of $\mathbb{B}$.
To conclude the proof of the upper bound in Theorem~\ref{sec-main} we need a
suitable upper bound for
\begin{equation}
\label{eq:rworigin}
\mathbb{E}^0\int\exp\Bigl\{-\int_0^t\beta(X_s)\,ds\Bigr\}\,\mathbb{B}(d\beta).
\end{equation}
To proceed, we consider the discrete-time analogue
$(\Tilde{X}_n)_{n\in\mathbb{N}}$ of $(X_t)_{t\ge0}$.
Let $(N_t)_{t\ge0}$ be the Poisson process with unit rate,
\begin{equation}
\Tilde{l}_n(x)=\sum_{m=0}^n\delta(x,\Tilde{X}_m)
\quad
\text{and}
\quad
\Tilde{R}_n=\{x\in\mathbb{Z}^d:\Tilde{X}_m=x \text{ for some } m=0,\ldots,n\}.
\end{equation}
The quantity in \eqref{eq:rworigin} equals
\begin{equation}
\label{eq:discrete}
\sum_{n=0}^\infty\mathbb{P}(N_t=n)\,
\Tilde{\mathbb{E}}^0\!\prod_{x\in\Tilde{R}_n}
\biggl(\int\frac{1}{1+\beta(0)}\,\mathbb{B}(d\beta)\biggr)^{\Tilde l_n(x)}
\end{equation}
Using $\Tilde l_n(x)\ge1$ when $x\in\Tilde R_n$ shows that \eqref{eq:discrete}
is bounded above by
\begin{equation}
\sum_{n=0}^\infty\mathbb{P}(N_t=n)\,\Tilde{\mathbb{E}}^0
\biggl(\int\frac{1}{1+\beta(0)}\,\mathbb{B}(d\beta)\biggr)^{|\Tilde{R}_n|}
\end{equation}
Returning to the continuous-time random walk $(X_t)_{t\ge0}$ we obtain
\begin{equation}
\mathbb{E}^0\int\exp\Bigl\{-\int_0^t\beta(X_s)\,ds\Bigr\}\,\mathbb{B}(d\beta)
\le
\mathbb{E}^0\biggl(\int\frac{1}{1+\beta(0)}\,\mathbb{B}(d\beta)\biggr)^{|R_t|}
\end{equation}
with $R_t$ the range of the continuous-time random walk $(X_t)_{t\ge0}$ in the
time interval $[0,t]$, i.e.,
\begin{equation}
\label{eq:rangerw}
R_t=\{x\in\mathbb{Z}^d:X_s=x \text{ for some } 0\le s\le t\}.
\end{equation}
Defining $\nu_1$ as in Remark~\ref{sec-constants}
and applying Theorem~\ref{sec-dovar}, completes the proof of the first
statement in Theorem~\ref{sec-main}.
\qed

\subsection{Lower bound}

The main purpose of the second statement of Theorem~\ref{sec-main} is to prove
that the upper bound in the first statement is optimal.
In the proof of the second part of Theorem~\ref{sec-main} we make use of the
following two lemmas.

\begin{lemma}
\label{sec-lemma1}
A local function $f:\mathbb{X}\rightarrow\mathbb{R}$ is monotone if and only
if
\begin{equation}
\sum_{\substack{A\subset B_2
                \\
                A\cap(B_2\setminus B_1)\not=\emptyset}}
\!\!\!
\Hat{f}(A)
\ge
0
\quad
\text{ for every }
B_1\subset B_2\subset\mathbb{Z}^d.
\end{equation}
\end{lemma}
\begin{proof}
The conclusion of the lemma follows from the observation
\begin{equation}
f(\zeta)-f(\eta)=\sum_{A\not=\emptyset}\Hat{f}(A)[H(\zeta,A)-H(\eta,A)]
\end{equation}
and the monotonicity of $H(\cdot,A)$ for every $A\in\mathbb{Y}$, because
$\eta\le\zeta$ if and only if there exist two subsets of $\mathbb{Z}^d$,
$B_1$ and $B_2$, such that $B_1\subset B_2$, $\eta=1$ on $B_1$, $\eta=0$ off
$B_1$, $\zeta=1$ on $B_2$, and $\zeta=0$ off $B_2$.
\end{proof}

\begin{lemma}
\label{sec-lemma2}
Suppose $\Lambda$ is a finite set, $\{x_A,A\subset\Lambda\}\subset\mathbb{R}$,
$\{y_A,A\subset\Lambda\}\subset[0,1]$, $x_\emptyset=0$,
$\sum_{B\subset A}x_B\ge0$ and $y_A=\prod_{a\in A}y_{\{a\}}$ for every
$A\subset\Lambda$, where it is understood that $y_\emptyset=1$.
Then, for\/ $|\Lambda|=2,3,\ldots$,
\begin{equation}
\label{eq:ineq1}
\sum_{A\subset\Lambda}x_Ay_A(1-y_{\Lambda\setminus A})
\ge
\sum_{a\in\Lambda}x_{\{a\}}y_{\{a\}}\prod_{b\in\Lambda\setminus\{a\}}(1-y_{\{b\}})
\ge
0.
\end{equation}
In particular,
\begin{equation}
\label{eq:ineq2}
\sum_{A\subset\Lambda}x_Ay_A\ge\Bigl(\sum_{A\subset\Lambda}x_A\Bigr)y_\Lambda.
\end{equation}
The last inequality also holds if\/ $|\Lambda|=1$.
\end{lemma}
\begin{proof}
Let, for $A\subset\Lambda$,
\begin{equation}
\label{eq:z(x)}
z_A=\sum_{B\subset A}x_B.
\end{equation}
Inverting \eqref{eq:z(x)} yields
\begin{equation}
\begin{split}
\sum_{A\subset\Lambda}x_Ay_A(1-y_{\Lambda\setminus A})
&=
\sum_{A\subset\Lambda}\Bigl(\sum_{B\subset A}(-1)^{|A\setminus B|}z_B\Bigr)
y_A(1-y_{\Lambda\setminus A})
\\
&=
\sum_{B\subset\Lambda}z_B\!\sum_{A:B\subset A\subset\Lambda}
(-1)^{|A\setminus B|}y_A(1-y_{\Lambda\setminus A})
\\
&=
\sum_{B\subset\Lambda}z_B\sum_{D\subset\Lambda\setminus B}(-1)^{|D|}
(y_{B\cup D}-y_{(\Lambda\setminus B)\cup B})
\\
&=
\sum_{B\subset\Lambda}z_By_B\sum_{D\subset\Lambda\setminus B}(-1)^{|D|}
(y_D-y_{\Lambda\setminus B})
\\
&=
\sum_{\substack{B\subset\Lambda\\ B\not=\emptyset}}
z_By_B\prod_{a\in\Lambda\setminus B}(1-y_{\{a\}})
\end{split}
\end{equation}
The last sum contains only non-negative terms. Thus, reducing it to a sum over
all singletons in $\Lambda$ gives a lower bound. Finally, remarking that
$z_{\{a\}}=x_{\{a\}}$ for every $a\in\Lambda$, proofs \eqref{eq:ineq1}.
\end{proof}

\noindent
Due to the attractiveness of the biased voter model, $S_\beta(t)f$ is
monotone whenever $f$ is monotone. In particular,
\begin{equation}
S_\beta(t)f-\delta_0(f)\ge0
\end{equation}
when $f$ is monotone
(because $\delta_0$ is invariant).
Moreover, applying inequality \eqref{eq:ineq2} of Lemma~\ref{sec-lemma2} to
$x_A=\Hat{f}(A)$ (see also Lemma~\ref{sec-lemma1}), $y_A=H(\eta_t,A)$, and
$\Lambda=\Lambda(f)$, shows that
\begin{equation}
S_\beta(t)f(\eta)-\delta_0(f)
=
\sum_{A\not=\emptyset}\Hat{f}(A)\mathbb{E}^\eta_\beta H(\eta_t,A)
\ge
\biggl(\sum_{A\not=\emptyset}\Hat{f}(A)\biggr)
\mathbb{E}^\eta_\beta H(\eta_t,\Lambda(f))\ge0
\end{equation}
Due to attractiveness and the monotonicity of $H(\cdot,A)$,
\begin{equation}
||S_\beta(t)f-\delta_0(f)||
\ge
\biggl(\sum_{A\not=\emptyset}\Hat{f}(A)\biggr)
\mathbb{E}^1_\beta H(\eta_t,\Lambda(f))
\end{equation}

\noindent
Now, we will derive a lower bound for
$\displaystyle\int\mathbb{E}^1_\beta H(\eta_t,\Lambda(f))\,\mathbb{B}(d\beta)$.
First notice that
\begin{equation}
\begin{split}
\int\mathbb{E}^1_\beta H(\eta_t,A)\,\mathbb{B}(d\beta)
&=
\int\mathbb{E}^A
\exp\biggl\{-\sum_{x\in\mathbb{Z}^d}\beta(x)l_t(x)\biggr\}
\,\mathbb{B}(d\beta)
\\
&=
\mathbb{E}^A\prod_{x\in\Bar R_t}
\int\exp\{-\beta(x)l_t(x)\}\,\mathbb{B}(d\beta)
\\
&\ge
\mathbb{E}^A\prod_{x\in\Bar R_t}\mathbb{B}(\{\beta:\beta(x)=0\})
\\
&=
\mathbb{E}^A\exp(-\nu_2|\Bar R_t|)
\end{split}
\end{equation}
where $\nu_2$ is defined in Remark~\ref{sec-constants},
\begin{equation}
\label{eq:rangedual}
l_t(x)=\int_0^t\mathbb{I}_{A_s}(x)\,ds
\quad
\text{and}
\quad
\Bar R_t=\!\!\bigcup_{0\le s\le t}\!\!A_s.
\end{equation}
If $A=\{x^1,\ldots,x^n\}$ with distict $x^i$'s, we couple
$(A_t)$ and $(X^1_t,\ldots,X^n_t)$ with $X^i_0=x^i$ in such a
way that $A_t\subset\{X^1_t,\ldots,X^n_t\}$ for all $t\ge0$, where
$X^1_t,\ldots,X^n_t$ are independent random walks with transition
probabilities $p_t(x)$ (identical to the coupling used in the proof of Lemma~1.5
of Chapter~V in Liggett (1985)).
Since in this coupling the range $\Bar R_t$ of the process $(A_t)$, see
\eqref{eq:rangedual}, can only be smaller than the range $R_t^n$ of the process
$(X^1_t,\ldots,X^n_t)$, i.e.,
\begin{equation}
R_t^n=\!\!\bigcup_{0\le s\le t}\!\!\{X_s^1,\ldots,X_s^n\},
\end{equation}
we find

\begin{equation}
\label{eq:coupling}
\begin{split}
\mathbb{E}^A\exp(-\nu_2|\Bar R_t|)
&\ge
\mathbb{E}^{(x^1,\ldots,x^n)}\exp(-\nu_2|R_t^n|)
\\[0.3 cm]
&\ge
\prod_{x\in A}\mathbb{E}^x\exp(-\nu_2|R_t|)
\\
&=
\bigl[\mathbb{E}^0\exp(-\nu_2|R_t|)\bigr]^{|A|}
\end{split}
\end{equation}
In the two last lines of \eqref{eq:coupling} $R_t$ is defined by
\eqref{eq:rangerw}.
Summarizing,
\begin{equation}
\log\int||S_\beta(t)f-\delta_0(f)||\,\mathbb{B}(d\beta)
\ge
\log\sum_{A\not=\emptyset}\Hat{f}(A)
+
|\Lambda(f)|\log\mathbb{E}^0\exp(-\nu_2|R_t|).
\end{equation}
Applying Theorem~\ref{sec-dovar} gives the desired result. \qed

\section*{Acknowledgements}

We thank Frank den Hollander for his cooperation in an early stage of this work.
SVG wishes to thank the ``Instituut voor Wetenschappelijk en Technologisch
onderzoek in de industrie (IWT)'' for financial support during a period when
part of this work was done.

\section*{References}

K. S. Alexander, F. Cesi, L. Chayes, C. Maes, and F. Martinelli,
{\em Convergence to equilibrium of random Ising models in the Griffiths' phase},
J. Stat.\ Phys.\ {\bf 92}, 337--352 (1998).
\\[0.2 cm]
A. J. Bray,
{\em Upper and lower bounds on dynamic correlations in the Griffiths phase},
J. Phys.\ A {\bf 22}, L81--L85 (1989).
\\[0.2 cm]
F. Cesi, C. Maes, and F. Martinelli,
{\em Relaxation of disordered magnets in the Griffiths' regime},
Comm.\ Math.\ Phys.\ {\bf 188}, 135--173 (1997a).
\\[0.2 cm]
F. Cesi, C. Maes, and F. Martinelli,
{\em Relaxation to equilibrium for two dimensional disordered Ising systems in
the Griffiths' phase},
Comm.\ Math.\ Phys.\ {\bf 189}, 323--335 (1997b).
\\[0.2 cm]
M. D. Donsker and S. R. S. Varadhan,
{\em On the number of distinct sites visited by a random walk},
Comm.\ Pure Appl.\ Math.\ {\bf 32}, 721--747 (1979).
\\[0.2 cm]
G. Gielis and C. Maes,
{\em Percolation techniques in disordered spin flip dynamics: relaxation to the
unique invariant measure},
Comm.\ Math.\ Phys.\ {\bf 177}, 83--101 (1996).
\\[0.2 cm]
A. Guionnet and B. Zegarlinski,
{\em Decay to equilibrium in random spin systems on a lattice},
Comm.\ Math.\ Phys.\ {\bf 181}, 703--732 (1996).
\\[0.2 cm]
A. Guionnet and B. Zegarlinski,
{\em Decay to equilibrium in random spin systems on a lattice II},
J. Stat.\ Phys.\ {\bf 86}, 899--904 (1997).
\\[0.2 cm]
W. Th.\ F. den Hollander,
{\em Random walks on lattices with randomly distributed traps. I. The average
number of steps until trapping},
J. Stat.\ Phys.\ {\bf 37}, 331--367 (1984).
\\[0.2 cm]
F. den Hollander, J. Naudts, and F. Redig,
{\em Invariance principle for the stochastic Lorentz lattice gas},
J. Stat.\ Phys.\ {\bf 66}, 1583--1598 (1992).
\\[0.2 cm]
F. den Hollander, J. Naudts, and F. Redig,
{\em Dynamic structure factor in a random diffusion model},
J. Stat.\ Phys.\ {\bf 76}, 1267--1286 (1994).
\\[0.2 cm]
F. den Hollander, J. Naudts, and P. Scheunders,
{\em A long-time tail for random walk in random scenery},
J. Stat.\ Phys.\ {\bf 66}, 1527--1555 (1992).
\\[0.2 cm]
F. den Hollander and K. E. Shuler,
{\em Random walks in a random field of decaying traps},
J. Stat.\ Phys.\ {\bf 67}, 13--31 (1992).
\\[0.2 cm]
A. Klein,
{\em Extinction of contact and percolation processes in a random environment\/},
Ann.\ Prob.\ {\bf 22}, 1227--1251 (1994).
\\[0.2 cm]
T. M. Liggett,
{\em Interacting Particle Systems\/}
\selectlanguage{german}
(Grundlehren der mathematischen Wissenschaften 273, Springer--Verlag, 1985).
\selectlanguage{english}
\\[0.2 cm]
F. Martinelli,
{\em Lectures on Glauber dynamics for discrete spin models\/}
(Proceedings of the Saint Flour summer school in probability 1997, to appear in
Lecture Notes in Mathematics, Springer--Verlag).
\\[0.2 cm]
A. T. Ogielski,
{\em Dynamics of three-dimensional Ising spin glasses in thermal equilibrium},
Phys.\ Rev.\ B {\bf 32}, 7384--7398 (1985).
\\[0.2 cm]
F. Redig,
{\em An exponential upper bound for the survival probability in a dynamic random
trap model},
J. Stat.\ Phys.\ {\bf 74}, 815--828 (1994).
\\[0.2 cm]
F. Spitzer,
{\em Principles of Random Walk\/}
(Graduate Texts in Mathematics 34, Springer--Verlag, 1976).
\\[0.2 cm]
B. Zegarlinski,
{\em Strong decay to equilibrium in one-dimensional random spin systems},
J. Stat.\ Phys.\ {\bf 77}, 717--732 (1994).
\end{document}